\documentclass[12pt]{article}
\usepackage{pdproc} 
\usepackage{epsfig} 
\usepackage{amsmath} 

\begin{document}

\renewcommand{\thefootnote}{\alph{footnote}}
  
\title{Counting Electrons to Probe the Neutrino Mass Hierarchy}

\author{J\"urgen Brunner}

\address{CPPM, Aix-Marseille Universit\'{e}, CNRS/IN2P3, Marseille, France}

\abstract{After the successful measurement of the mixing angle $\theta_{13}$, 
the determination of the neutrino mass hierarchy has become a priority 
for future neutrino experiments. We propose a conventional $\nu_\mu$ beam with neutrino
energies in the range 2-8 GeV aimed at a Mton underwater detector at the ``magic" baseline of 
2600~km. In this constellation it is sufficient to distinguish  ($\nu_\mu$ induced) track-like
interactions from cascade-like interactions with moderate purity to determine the mass hierarchy. 
}

\normalsize\baselineskip=15pt

\section{Introduction}
All three mixing angles and both mass square differences of the neutrino mass Eigenstates are known after
the measurement of the mixing angle $\theta_{13}$~\cite{daya,reno}.
Global fits of all experimental
input~\cite{fogli} provide a coherent picture of the oscillation parameters.
Among the yet unknown features of the oscillation scheme, the neutrino mass hierarchy (MH) 
is considered to be in reach for the next generation of experiments.
It has been proposed 
to determine the MH by measuring atmospheric neutrinos~\cite{ars} 
in planned low-energy extensions 
of existing neutrino telescopes~\cite{pingu,orca}. 
The distinction of the two hierarchy hypotheses with atmospheric neutrinos is challenging due to cancellation 
of contributions from
neutrinos and anti-neutrinos and the effect is further attenuated by the finite energy and 
angular resolutions of the considered detectors. Both problems are avoided by counting beam related events 
of a specific flavour.
Such a concept has been recently proposed~\cite{vissani}: muon
counting from $\nu_\mu$ interactions in very long baseline beams (L $>$ 6000~km) should allow
MH determination in Mton underwater/ice detectors. However, the construction of a steeply inclined beam-line
has never been performed and is technically challenging and costly. 
Here, we propose instead to point a beam with a baseline of about 2600~km to a Mton underwater detector
and to count $\nu_e$ interactions.

\section{Oscillation Probabilities}

Oscillation probabilities are calculated in a full three flavour scheme 
using the Globes package~\cite{globes}.
\begin{figure}[htpb]
\centering
\epsfig{figure=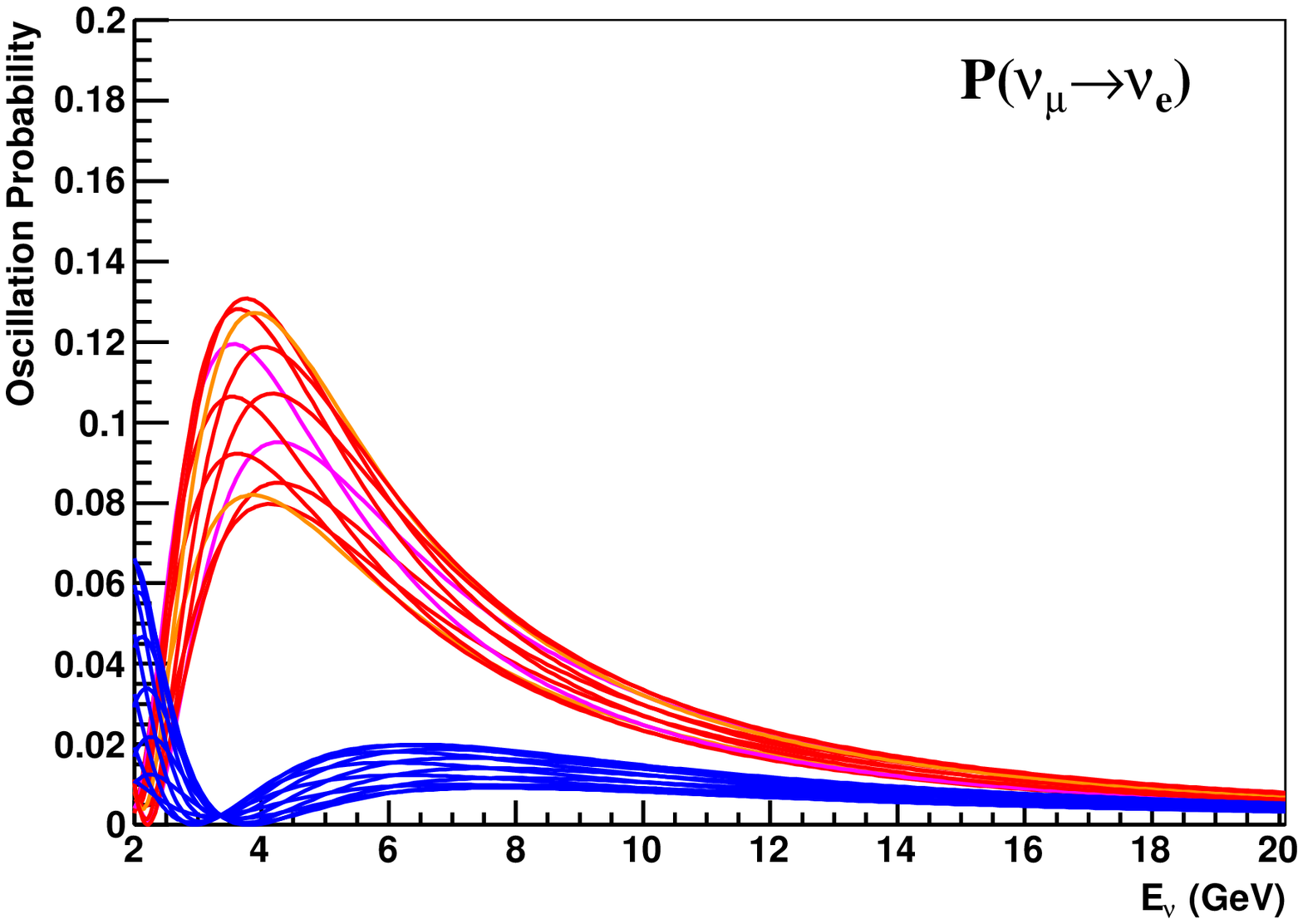,width=7.5cm}
\epsfig{figure=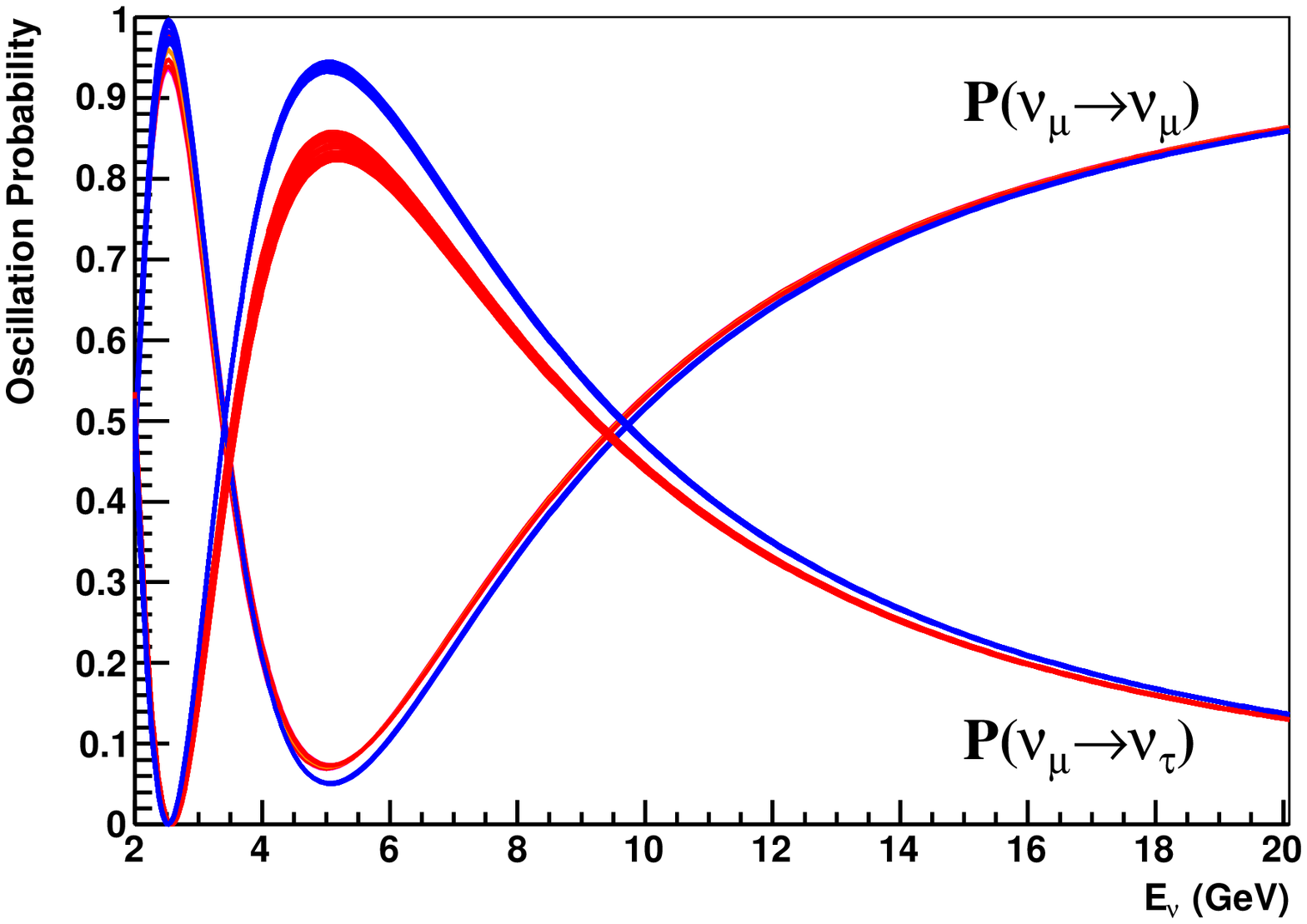,width=7.5cm}
\caption{
Neutrino oscillation probabilities $P(\nu_\mu\rightarrow\nu_e)$,
$P(\nu_\mu\rightarrow\nu_\mu)$ and $P(\nu_\mu\rightarrow\nu_\tau)$
for a baseline of 2600~km and oscillation
parameters from a global fit as a function of
neutrino energy. The group of red lines is for NH, blue for IH, $\phi_{CP}$ is varied in steps
of 30$^\circ$ between $0^\circ$ and 330$^\circ$. The special values $\phi_{CP}=0^\circ,180^\circ$ (CP conservation) 
are indicated in magenta, $\phi_{CP}=90^\circ,270^\circ$ (maximal CP violation) in orange on the left plot.
}
\label{posc}
\end{figure}
For a given set of neutrino parameters for normal mass hierarchy (NH),
a convention is needed to transform them into inverted hierarchy (IH). We follow the approach proposed in Ref.~\cite{fogli}, 
defining an average ``large" mass square difference $\Delta M^2$ which controls atmospheric neutrino oscillations
\begin{equation}
\Delta M^2 = \frac{1}{2}\left( \Delta m^2_{32,NH} + \Delta m^2_{31,NH}\right)
= \frac{1}{2}\left( \Delta m^2_{23,IH} + \Delta m^2_{13,IH}\right)
\end{equation}
with $\Delta m^2_{ab}=m^2_a - m^2_b$.
IH is defined as the sign change of $\Delta M^2$ for a given set of parameters in NH. 
This leads to the following transformation rules for the related mass square differences:
\begin{eqnarray}
\Delta m^2_{23,IH} &= \Delta m^2_{32,NH} + \Delta m^2_{21} \\
\Delta m^2_{13,IH} &= \Delta m^2_{31,NH} - \Delta m^2_{21}
\end{eqnarray}
Figure~\ref{posc} shows the resulting oscillation probabilities $P(\nu_\mu\rightarrow\nu_e,\nu_\mu,\nu_\tau)$ with the best fit parameters
from Ref.~\cite{fogli} for a baseline of 2600~km. Most of the neutrino path will be in the outer Earth mantle with a maximal depth of 134~km.
A constant density of 3.3~g/cm$^3$ is used for the calculation of the oscillation probabilities as given in the 
PREM model~\cite{prem} for the outer mantle. 

For $P(\nu_\mu\rightarrow\nu_e)$ a significant difference between both hierarchies is observed. 
The variation of the CP-phase $\phi_{CP}$ (different lines of the same color) leads instead only to moderate changes 
of the oscillation probabilities.
For $E_\nu=3.5$~GeV and IH $P(\nu_\mu\rightarrow\nu_e)$ is strictly independent from $\phi_{CP}$. 
In the range 3~GeV$ < E_\nu < $8~GeV there is no overlap of the two CP-bands for the two hierarchies which allows to determine MH by counting $\nu_e$ events.
This special feature has been noted in the past~\cite{magic} and led to the label ``magic" for baselines in the range 2500-2600~km.
The large value of the mixing angle $\theta_{13}\approx 9^\circ$ leads to peak values of 13\% for $P(\nu_\mu\rightarrow\nu_e)$ in NH, allowing to detect
a sizable sample of $\nu_e$ events in a suitable detector.

$P(\nu_\mu\rightarrow\nu_\mu)$ depends only weakly on MH and not on $\phi_{CP}$. The first vacuum oscillation minimum at 5~GeV is the dominating feature in
the shown energy range. Counting $\nu_\mu$ events can serve as a flux normalisation or it could be used to improve the measurement 
of the atmospheric oscillation parameters
$\Delta M^2$ and $\theta_{23}$.

\section{Cross Sections}

\begin{figure}[htpb]
\centering
      \epsfig{figure=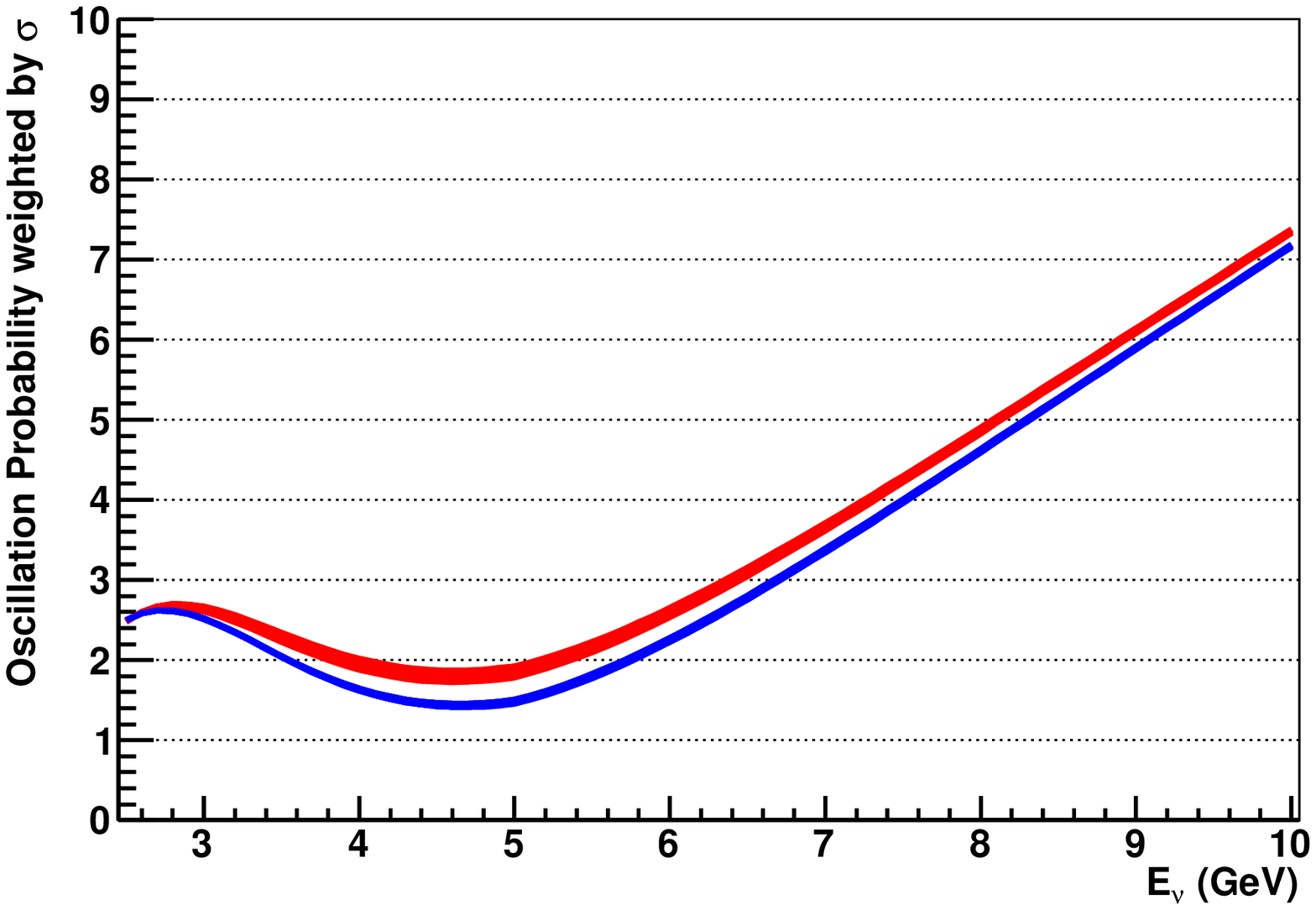,width=7.5cm}
      \epsfig{figure=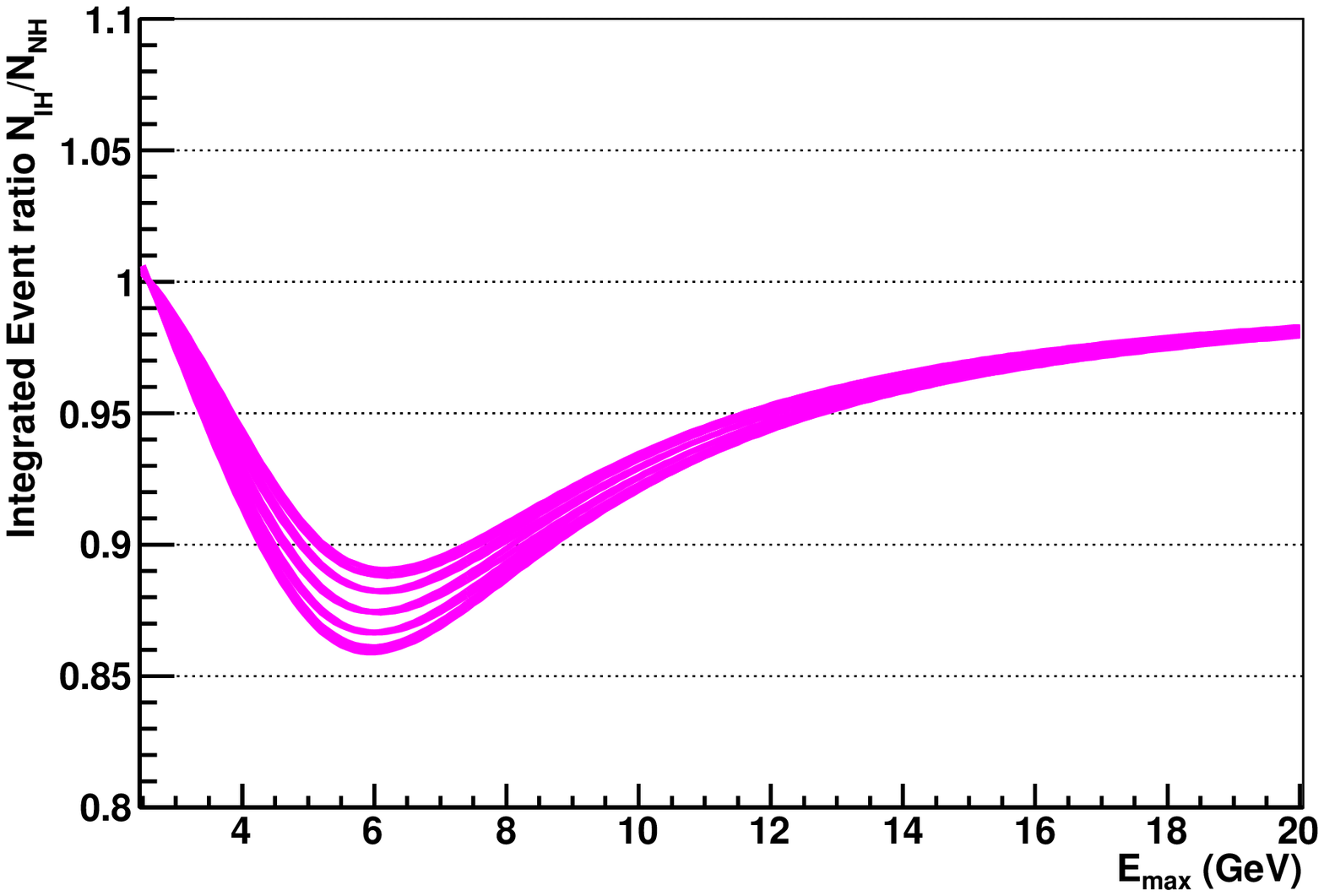,width=7.5cm}
\caption{
{\bf Left:} Summed oscillation probabilities $P(\mu\rightarrow\alpha)$ weighted by cross sections (red : NH, blue : IH)
{\bf Right:} Ratio of Integrals $N_{IH}/N_{NH}$ from left plot between $E_{min}=$2.5~GeV and $E_{max}$.
}\label{sigweight}
\end{figure}

The total cross sections for $\nu_\mu$ and $\bar\nu_\mu$ charged current (CC) interactions are taken from Ref.~\cite{pdg} in the parton scaling approximation:
\begin{eqnarray}
&\sigma^{CC}_{\nu_\mu}(E_{\nu}) &= 0.68\cdot (E_\nu/GeV) 10^{-38} \mbox{cm}^2 \\
&\sigma^{CC}_{\bar{\nu}_\mu}(E_\nu) &= 0.34\cdot (E_\nu/GeV) 10^{-38} \mbox{cm}^2.
\end{eqnarray}
Deviations from this linear behaviour due to quasi-elastic or resonant interactions are ignored. Their contribution would not alter
the result of this study in a significant way.
The relevant neutrino energies are significantly larger than $m_\mu$ and $m_e$ therefore from flavour universality
$\sigma^{CC}_{\nu_e}=\sigma^{CC}_{\nu_\mu}$ and 
$\sigma^{CC}_{\bar\nu_e}=\sigma^{CC}_{\bar\nu_\mu}$.
However for $\nu_\tau$ CC interactions the mass of the $\tau$-lepton cannot be neglected. We use the calculation from Ref.~\cite{nutau}.
For a threshold energy $E_0=5$~GeV and $E_0 < E_\nu <$~30~GeV we find the following simple parametrisation:
\begin{equation}
\sigma^{CC}_{\nu_\tau}(E_\nu) = 0.29\log\left(\frac{E_\nu}{E_0}\right)\sigma^{CC}_{\nu_\mu}(E_\nu).
\end{equation}
The cross sections for neutral current (NC) interactions for all flavours are approximated as:
\begin{eqnarray}
&\sigma^{NC}_\nu(E_\nu) &= \frac{1}{3}\sigma^{CC}_{\nu_\mu}(E_{\nu})\\
&\sigma^{NC}_{\bar{\nu}}(E_\nu) &= \frac{1}{3}\sigma^{CC}_{\bar{\nu}_\mu}(E_\nu).
\end{eqnarray}
Whereas the sum of the oscillation probabilities $\sum_\alpha P(\mu\rightarrow\alpha)$
equals unity due to the unitarity of the flavour mixing matrix, the cross section weighted sum
\begin{equation}
P_\mu^\sigma(E_\nu) = \frac{\sigma^{NC}_\nu(E_\nu)+
\sum_\alpha P(\mu\rightarrow\alpha)\sigma^{CC}_{\nu_\alpha}(E_\nu)}
{\sigma^{NC}_\nu(E_0)+\sigma^{CC}_{\nu_\mu}(E_0)}
\end{equation}
(arbitrarily normalised at $E_0 = 1$~GeV) may help identifying the optimal energy range to separate the two MH hypotheses.
$P_\mu^\sigma$ is shown on the left plot of Figure~\ref{sigweight}
and can be interpreted as the event rate per neutrino energy seen in a detector 
with an energy independent detection efficiency for a pure $\nu_\mu$ flux flat in energy. 
No distinction is made neither between NC and CC interactions nor between different flavours.
Nevertheless a clear separation of the two MH hypotheses is observed for energies above 3~GeV. 
It can be attributed to the 
kinematical suppression of $\nu_\tau$ CC interactions. The size of the effect is quantified in the right plot of Figure~\ref{sigweight} 
which shows the ratio between IH and NH integrals from the left plot (for different values of $\phi_{CP}$) taken between an assumed threshold 
energy of 2.5~GeV and a variable maximal energy $E_{max}$. An optimal value of
$E_{max}=6$~GeV can be read from the figure, which yields a suppression of the IH event
rate of 11-14\% compared to the expected rate for NH. Extending the energy range to higher values reduces the relative size of the
separation of the MH hypotheses. 
It can be concluded that a measurement of the mass hierarchy will be possible even without any flavour tagging capabilities using a
neutrino beam in a limited energy range of 2-6~GeV and a large detector which can reliably count beam related neutrino interactions.
Nevertheless flavour tagging methods are discussed below and they are used to improve the significance of the measurement.

\section{Neutrino beam}

The Institute of High Energy Physics (IHEP) in Protvino near Moscow (located at $54^\circ52' N, 37^\circ11' E$~\cite{maps})
hosts the U70 proton accelerator~\cite{skat} which provides protons with energies up to 70~GeV.
It is operational since 1967 and had been the world largest proton accelerator at its time of commissioning.

To obtain a powerful neutrino flux, a high intensity proton beam is needed.
A scheduled upgrade of the Fermilab accelerator complex is foreseen to yield $N_{pot}=3.6\cdot 10^{21}$ 
protons on target for the NOVA experiment within 6 years of operation~\cite{pot}. 
We assume that a similar performance can be reached with the U70 accelerator after a corresponding upgrade.
In the following, event numbers are calculated for $N_{pot}=1.5\cdot 10^{21}$ 
which might be achievable within 3-5 years of operation after such an upgrade.
\begin{figure}[htpb]
\centering
\epsfig{figure=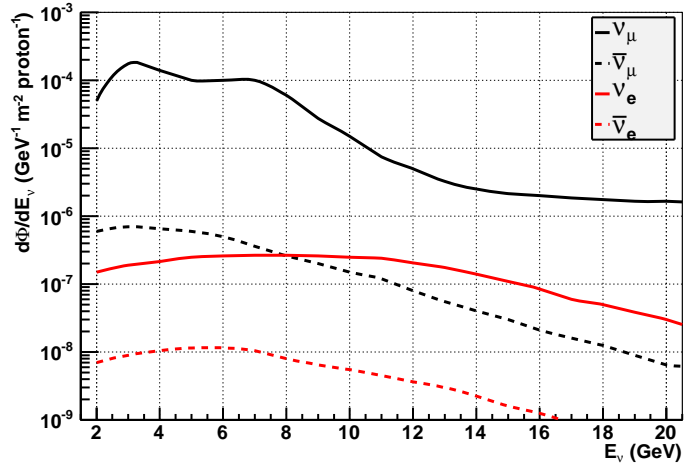,width=10.0cm}
\caption{Neutrino Flux $d\Phi_\nu/dE_\nu$ per proton as seen by the SKAT bubble chamber. 
The different lines correspond to the different neutrino flavours.
}\label{skat}
\end{figure}

In the past, a neutrino beam was provided to several experiments, 
among them the bubble chamber SKAT~\cite{skat}. Secondary hadrons were produced in an Aluminum target and focused by
parabolic lenses. Neutrinos were produced in a 140~m long decay line. The bubble chamber was situated 270~m behind the
target, and at a distance of $l_{SKAT}=245$~m downstream the beginning of the decay tube. 
This value will be used in the following to scale the beam intensity to the remote location.  
The neutrino fluxes $d\Phi_\nu/dE_\nu$, as they were delivered to the SKAT 
experiment for focusing of positively charged hadrons and for all four flavours present in the beam, are shown in Figure~\ref{skat}. 
A parametrisation of the beam intensity as function of the neutrino energy at the SKAT detector
has been obtained from Ref.~\cite{skat,skat1} and will be used in the following. 
As seen from Figure~\ref{skat}, a very clean $\nu_\mu$ beam was provided with energies dominantly in the range 2-8~GeV. 
Contaminations from other flavours were on the sub-percent level.

To provide a beam to a detector in the Mediterranean Sea at a distance of $l_{LBL}\approx$~2600~km a new beam-line is needed.
It would point in the southwest direction from the proton accelerator ring, 
an area which is currently not obstructed by buildings~\cite{maps}.
A moderate downward inclination of $11.7^\circ$ is needed. 

\section{Detector}
\label{sec:det}
The ANTARES detector~\cite{antares} is a deep sea neutrino telescope, which operates successfully in the Mediterranean Sea at 
($42^\circ48' N , 6^\circ10' E$). A neutrino beam from Protvino to this location would result in a baseline of 2588~km.
Recently, the ANTARES Collaboration published a measurement of atmospheric neutrino oscillations~\cite{antosc}, 
demonstrating the capability of the device to detect and measure neutrinos with energies as low as 20~GeV.
A multi-cubic-kilometer detector KM3NeT~\cite{km3net} is planned as a future neutrino telescope in the Mediterranean Sea. One possible site
would be close to the existing ANTARES detector. An alternative site in the Ionian Sea off the Sicilian coast results incidentally in 
an identical baseline (within 1\%) for a beam from Protvino. Therefore the calculation, presented here, holds for both site options.
\begin{figure}[htpb]
\centering
\epsfig{figure=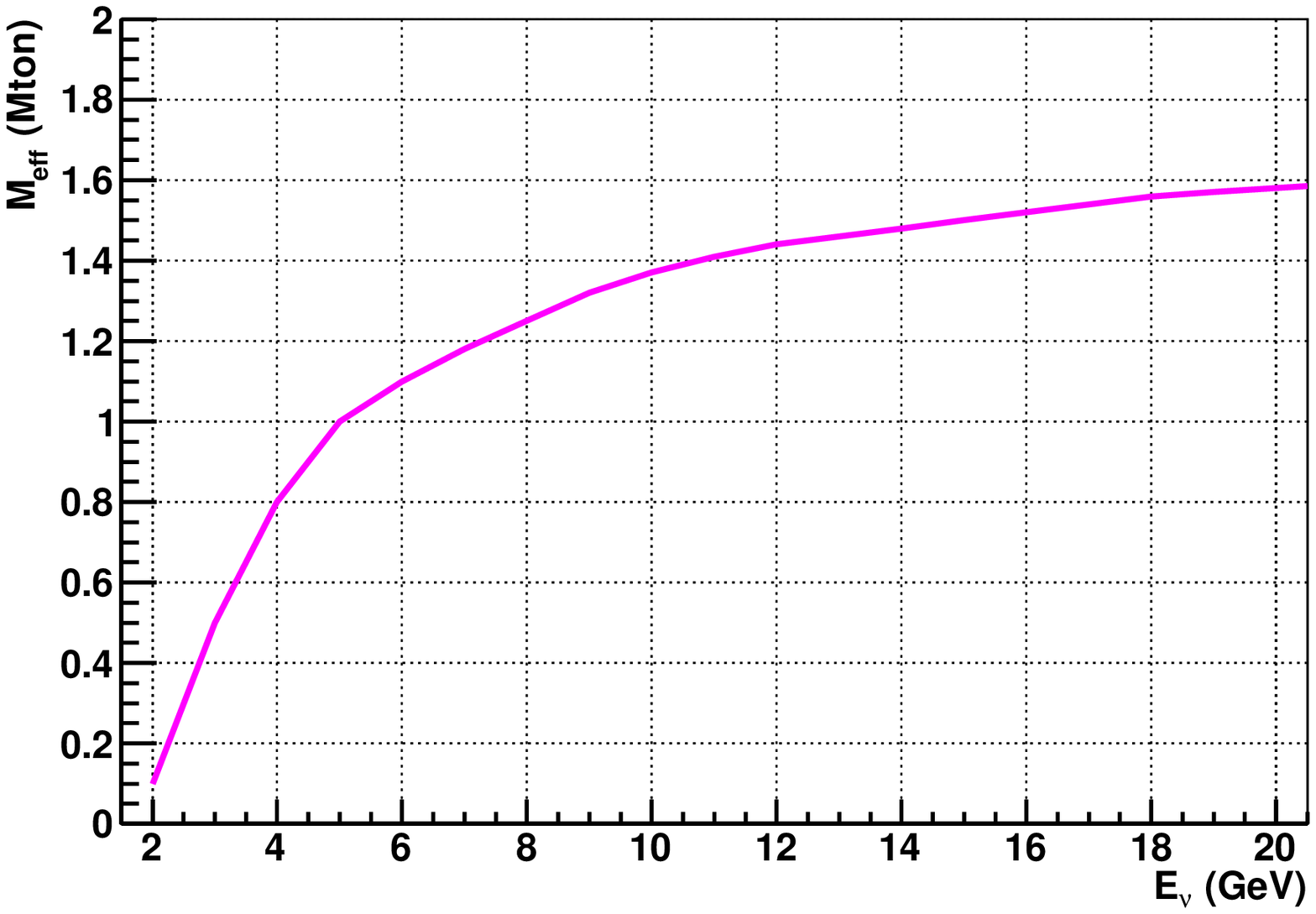,width=7.5cm}
\epsfig{figure=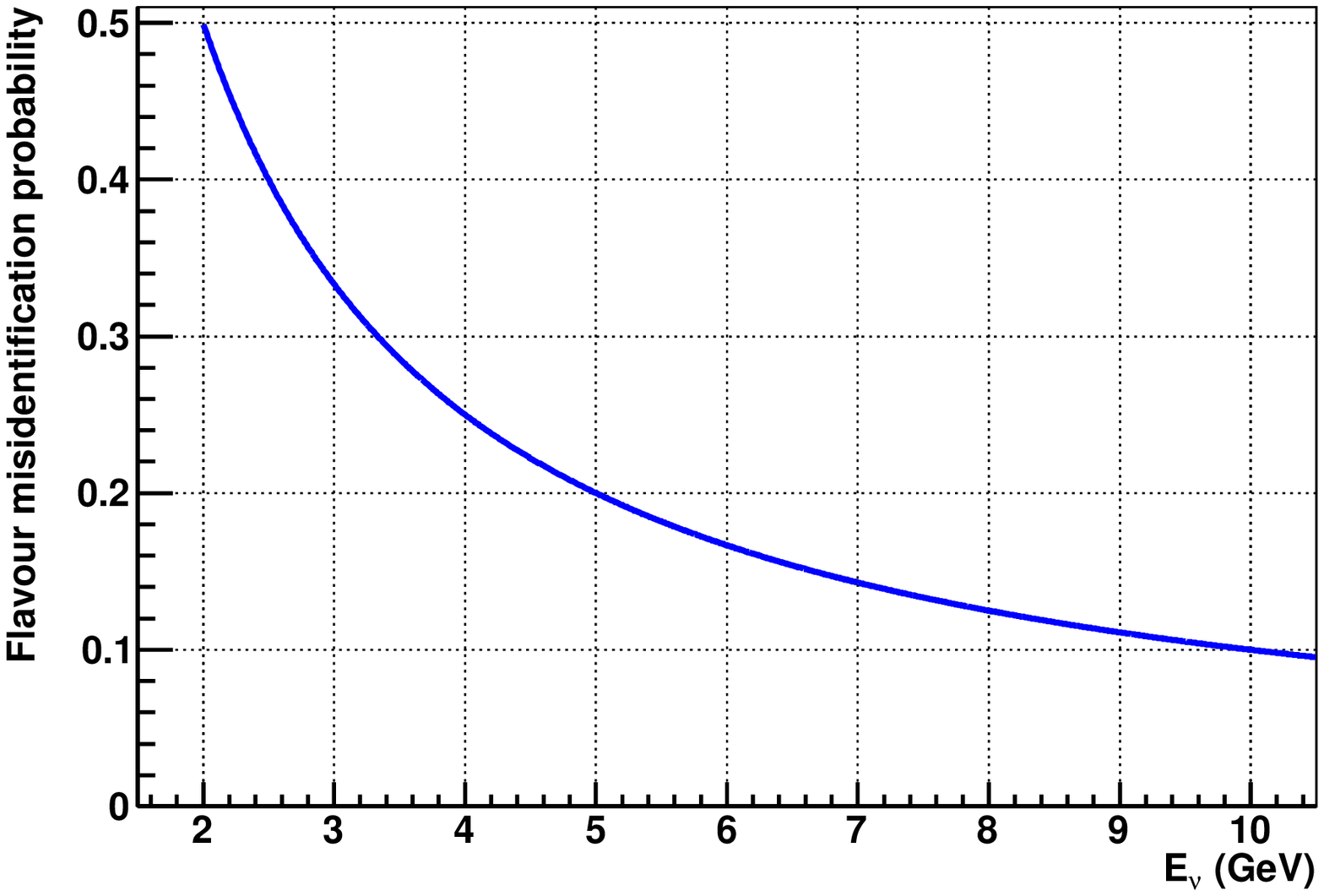,width=7.5cm}
\caption{{\bf Left:} Effective mass $M_{eff}$ for the detection of $\nu_\mu$ CC interactions as function of $E_\nu$ 
for events with the interaction vertex inside the instrumented volume. 
A successful track reconstruction including a condition on the likelihood of the track fit is required. 
{\bf Right: } Flavour misidentification probability as function of $E_\nu$. 
The same function is assumed for the probability to misidentify tracks as cascades and vice versa.
}\label{veff}
\end{figure}

Currently the KM3NeT Collaboration carries out a feasibility study to determine the physics reach for MH determination of a densely equipped 
detector based on about 20\% of the final budget~\cite{orca}. About 1000 optical modules would be used to instrument a water mass 
of around 2~Mtons. The detector is supposed to be sensitive to neutrinos with 
energies down to few GeV. Figure~\ref{veff} shows a preliminary result of this study~\cite{orca}:
the effective mass $M_{eff}$ for $\nu_\mu$ charged current interactions occurring inside the instrumented volume is given as function 
of the neutrino energy.
Here a trigger condition, a likelihood based track reconstruction and a quality criteria for the goodness of the track fit are applied. 

It is assumed that the considered detector is able to distinguish ``track" and ``cascade" event signatures. A ``track" requires the
presence of a long muon trajectory. As the muon range increases linearly with energy whereas the longitudinal extension 
of cascades remains essentially unchanged for the considered energies, 
the distinguishability of these two topologies will be a function of the neutrino energy.
For the event rate calculation below, we introduce $\epsilon(E_\nu)$ as the probability 
to misidentify a cascade as track, while the misidentification of 
a track as cascade will be called $\eta(E_\nu)$. Both functions are parametrised by
\begin{equation}
\epsilon(E_\nu) = \eta(E_\nu) = 1/(E_\nu/\mbox{GeV}) ; E_\nu > 2 \mbox{GeV}
\label{epseta}
\end{equation}
which is illustrated in the right plot of Figure~\ref{veff}. For $E_\nu = 2$~GeV, $\epsilon=\eta=0.5$, which means 
the two topologies cannot be distinguished and the attribution of an event to one of them is random.
A 5~GeV neutrino produces a muon with an average range of 15~m in Sea water which exceeds already by far the
typical longitudinal size of a hadronic or electromagnetic shower. 
Correspondingly the misidentification probability is assumed to drop to 20\%.
For $E_\nu = 10$~GeV $\epsilon$ and $\eta$ further decrease to 10\%. However, these low values of the 
misidentification probability are not exploited in this analysis due to the energy profile of
the beam, which suppresses contributions of neutrinos with $E_\nu > 10$~GeV.

\section{Signal Event Rates}

The rate of detected CC events of flavour $\alpha$ can now be calculated:
\begin{equation}
\begin{split}
\frac{dN_{\alpha}}{dE_\nu} = 
&N_{pot}\left(\frac{l_{SKAT}}{l_{LBL}}\right)^2\frac{M_{eff}(E_\nu)}{m_p} \\ 
&\left[\sigma^{CC}_{\nu_\alpha}\left(\frac{d\Phi_{\nu_\mu}}{dE_\nu}P_{\mu\alpha} 
+ \frac{d\Phi_{\nu_e}}{dE_\nu}P_{e\alpha}\right)
+\sigma^{CC}_{\bar{\nu}_\alpha}\left(\frac{d\Phi_{\bar{\nu}_\mu}}{dE_\nu}P_{\overline{\mu\alpha}} 
+ \frac{d\Phi_{\bar{\nu}_e}}{dE_\nu}P_{\overline{e\alpha}}\right)\right]
\end{split}
\label{cc}
\end{equation}
with $m_p$ the proton mass and the abbreviation $P_{\beta\alpha} = P(\nu_\beta\rightarrow\nu_\alpha)$ 
and $P_{\overline{\beta\alpha}} = P(\bar\nu_\beta\rightarrow\bar\nu_\alpha)$ for the oscillation probabilities. 
All four initial-state neutrino flavours which are present in the neutrino beam are taken into account.
The major contribution for all final-state flavours come from the dominant $\nu_\mu$-beam component.

Information on reconstruction efficiencies for $\nu_e$ and $\nu_\tau$ CC events are not available yet.
Contained $\nu_e$ and $\nu_\mu$ CC events will on average deposit the same amount of energy in the detector,
therefore the same neutrino energy dependent effective mass $M_{eff}(E_\nu)$ from Figure~\ref{veff}
is assumed for both. The same efficiency function is also used for $\nu_\tau$ CC events, 
despite the fact that they
release less energy in the detector due to the escaping neutrino(s) from the tau decay.
As $\nu_\tau$ interactions are a background in the present analysis, this is a conservative approximation.

Similarly the rate of detected NC events can be calculated
\begin{equation}
\begin{split}
\frac{dN_{NC}}{dE_\nu} = 
&N_{pot}\left(\frac{l_{SKAT}}{l_{LBL}}\right)^2\frac{M_{eff}(E_\nu/2)}{m_p}  \\
&\left[\sigma^{NC}_\nu\left(\frac{d\Phi_{\nu_\mu}}{dE_\nu} + \frac{d\Phi_{\nu_e}}{dE_\nu}\right)
+\sigma^{NC}_{\bar{\nu}}\left(\frac{d\Phi_{\bar{\nu}_\mu}}{dE_\nu} 
+ \frac{d\Phi_{\bar{\nu}_e}}{dE_\nu}\right)\right].
\end{split}
\label{nc}
\end{equation}
No dependence on oscillation parameters enters here. As on average 50\% of the neutrino energy is transferred to
the outgoing neutrino, the effective mass $M_{eff}$ is evaluated at $E_\nu/2$.
\begin{figure}[htpb]
\centering
      \epsfig{figure=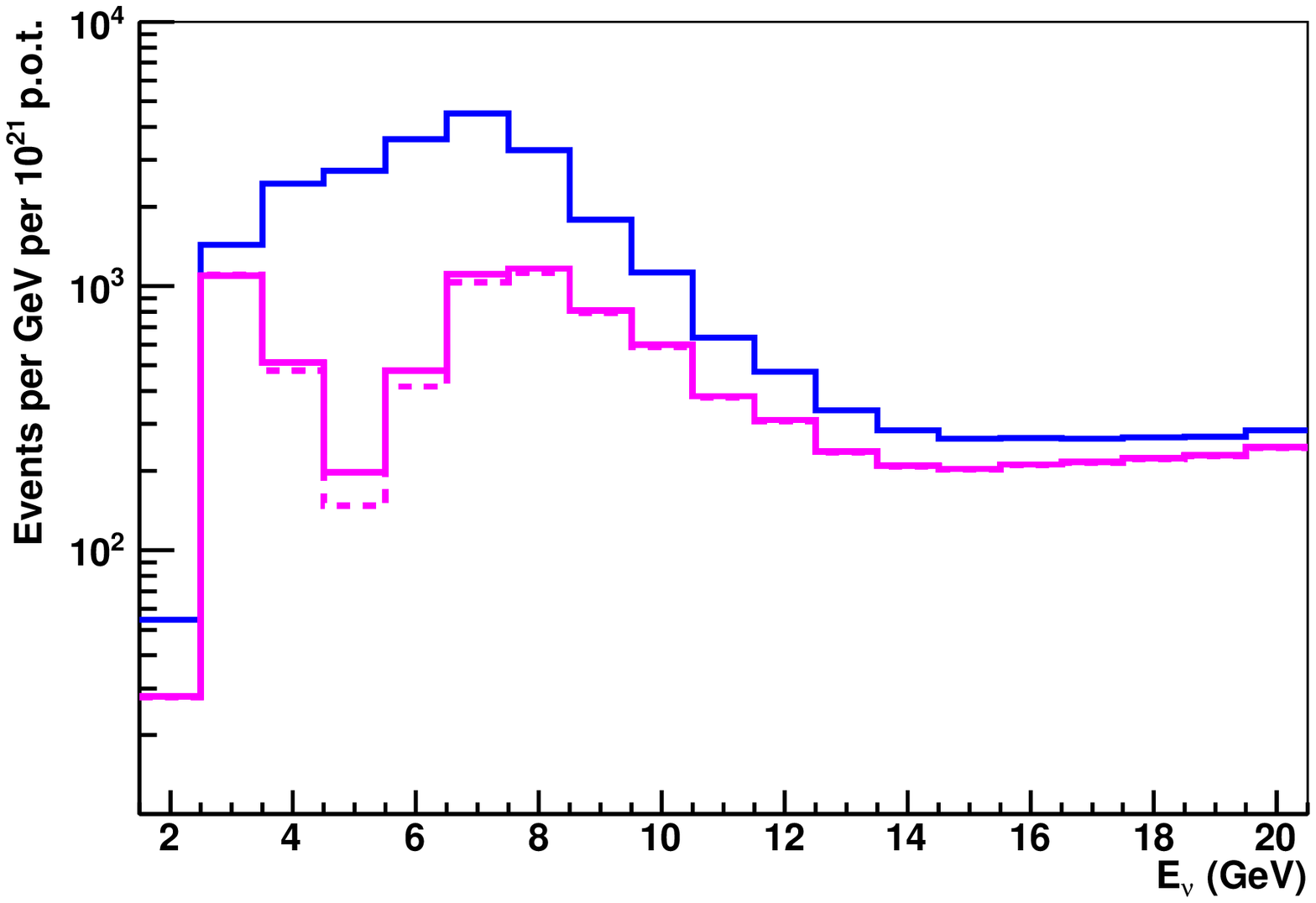,width=7.5cm}
      \epsfig{figure=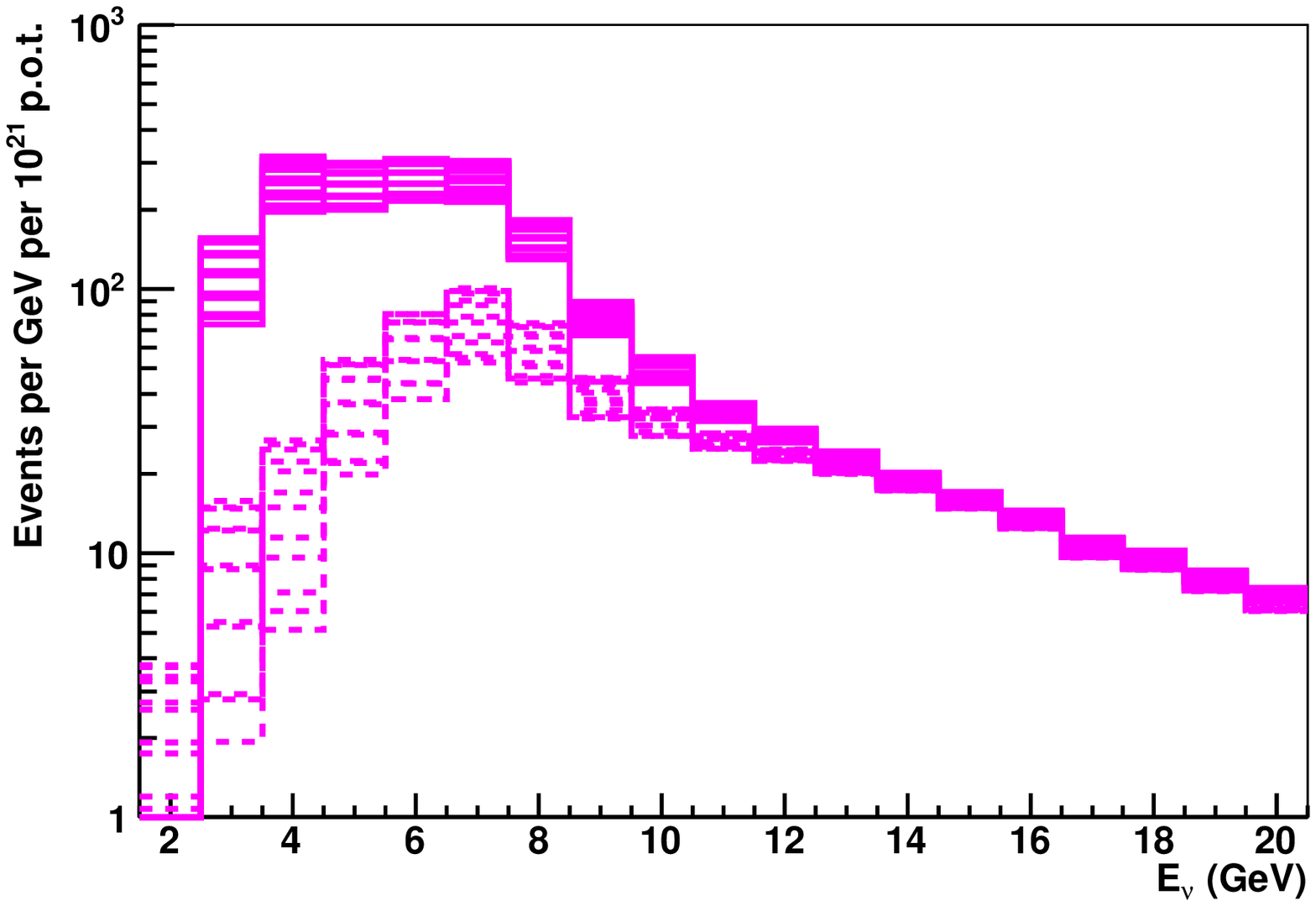,width=7.5cm}
\caption{Observed events for $N_{pot}=1.5\cdot10^{21}$ as function of the neutrino energy.
{\bf Left:} Track-like events from $\nu_\mu$ CC interactions. {\bf Right:} Cascade-Like events
from $\nu_e$ CC interactions. 
The solid blue histogram on the left plot is obtained by ignoring neutrino oscillations.
Solid magenta histograms are for NH and various values of $\phi_{CP}$, dashed lines indicate IH.
}\label{clean}
\end{figure}

The ``track-signal" events are then simply given by
$dN_{sig}^{track}/dE_\nu = dN_{\mu}/dE_\nu$,
whereas the ``cascade-signal" will be defined as
$dN_{sig}^{casc}/dE_\nu = dN_{e}/dE_\nu$.
$\nu_\tau$ CC and NC interactions are considered as background.
Figure~\ref{clean} shows the expected signal event rates. The corresponding integral event numbers are summarized
in Table~\ref{tabclean}. At this stage we do not consider the misidentification between the two event topologies, 
{\it i.e.} $\epsilon = \eta = 0$.
The $\nu_\mu$ CC rate is suppressed by 60\% due to neutrino oscillations
compared to the no-oscillation hypothesis. 
Nonetheless a comfortable event sample of 10000~events can be detected for $N_{pot}=1.5\cdot10^{21}$ 
with a statistical uncertainty of 1\% whereas the two mass hierarchy hypotheses modify the
expected event rate by 3\%.
\begin{table}[htpb]
\begin{center}
\begin{tabular}{||l|c|c|c|c||} \hline
Channel & Tracks NH & Tracks IH & Cascades NH & Cascades IH \\ \hline 
No oscil          & \multicolumn{2}{c|}{26315}     &  \multicolumn{2}{c||}{---} \\ \hline
Signal         &  10317  &  10015     &  1366-1876 & 397-597 \\ \hline
$\nu_\tau$        &  227-231  & 245-248  &  1076-1098 & 1163-1176  \\
NC                & 0 & 0   & 4732 & 4732   \\ \hline
BG Total       & 227-231 & 245-248 &    5807-5830 & 5895-5908 \\ \hline
Total &  10543-10548 & 10260-10263  & 7196-7683 & 6304-6492\\ \hline
\end{tabular}
\end{center}
\caption{Event numbers for $N_{pot}=1.5\cdot10^{21}$ in the track and cascade channel for both mass hierarchy schemes 
and varying $\phi_{CP}$ values with a perfect separation of track and cascade signatures.}
\label{tabclean}
\end{table}

The right plot of Figure~\ref{clean} shows the event rate of $\nu_e$ CC events. For NH $1621\pm255$ events are
expected, where the uncertainty is due to the unknown CP-phase. This has to be compared to $497\pm100$ events for IH.
The statistical separation of both samples is better than 20$\sigma$. The largest effect is seen for neutrino
energies from 3-8~GeV, as expected. Despite the fact that Figure~\ref{clean} illustrates the $E_\nu$ distribution
of the selected event samples, no assumption about the energy determination is needed here to determine MH. 
However, a moderate energy resolution will certainly increase the significance of the MH hypothesis test. 

\section{Background and Purity of Event Selection}

The measurement will be complicated by background. Contributions from atmospheric neutrinos and misreconstructed
down-going atmospheric muons can be ignored. A pulsed beam with a typical duty cycle lower than $10^{-6}$
allows to safely discard these events. The beam itself is a source of background events. 
$\nu_\tau$ CC events with a muonic $\tau$-decay produce ``track" events, whereas 
NC events and $\nu_\tau$ CC events with a non-muonic $\tau$-decay have a genuine cascade signature. 
These two contributions are added in Table~\ref{tabclean} and they are the only background contributions, if we
assume a perfect flavour tagging mechanism to separate track-like from cascade-like events.
Whereas the sample of track-like events is only marginally affected by the addition of $\nu_\tau$ events,
the cascade event sample is now dominated by NC events which contribute about three times as much as 
the $\nu_e$ signal events.  

\begin{figure}[htpb]
\centering
      \epsfig{figure=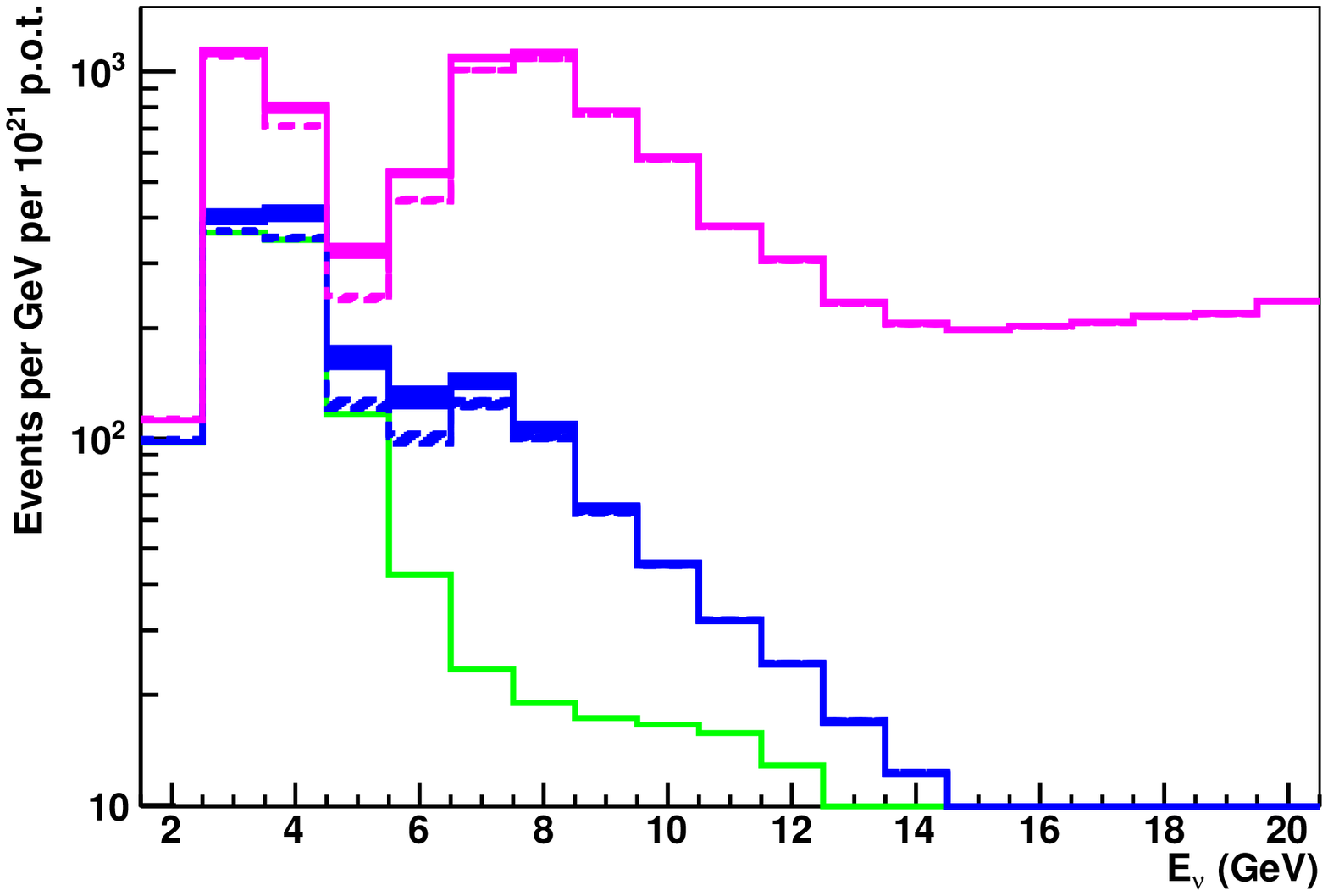,width=7.5cm}
      \epsfig{figure=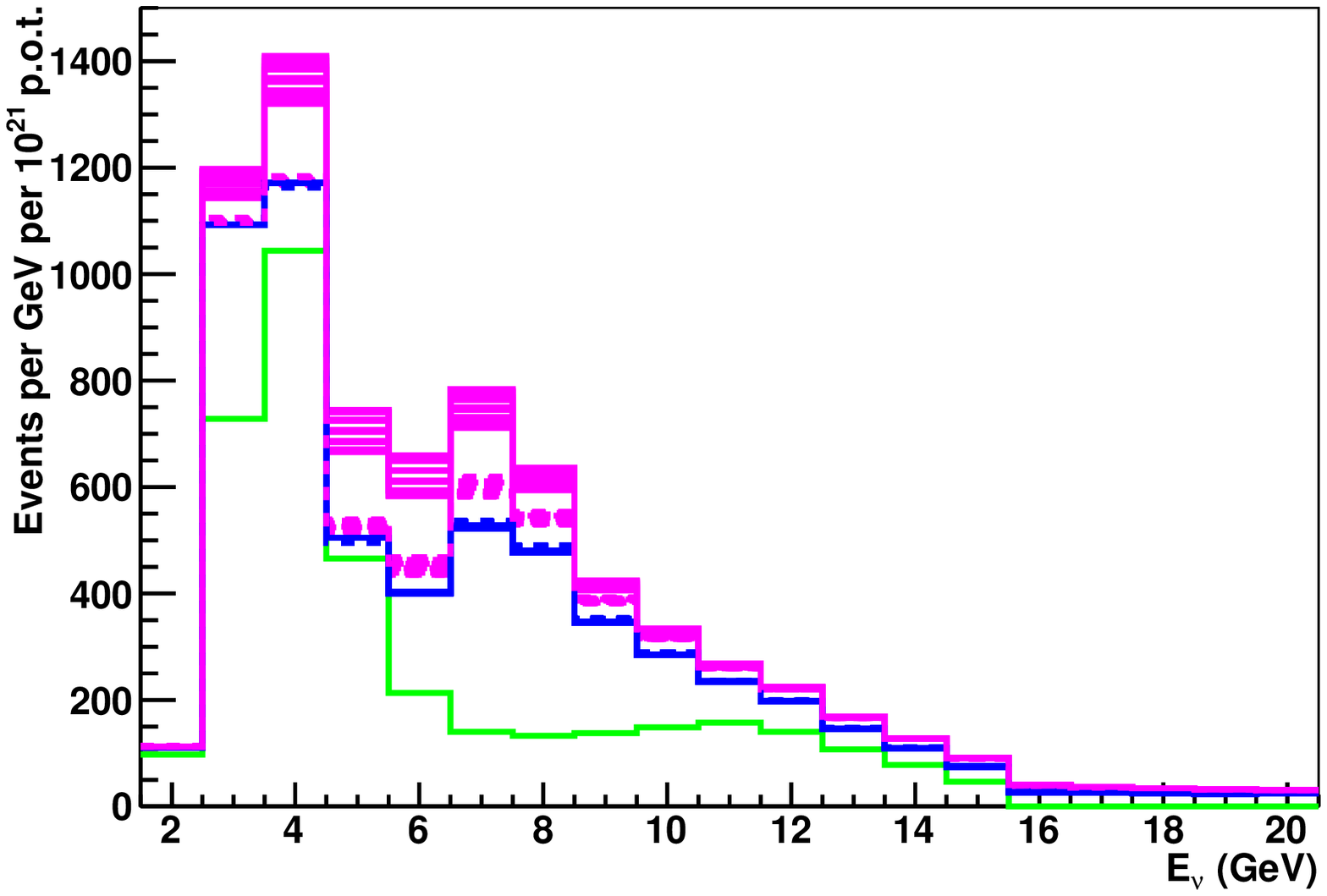,width=7.5cm}
\caption{Observed events for $N_{pot}=1.5\cdot10^{21}$ as function of the neutrino energy. 
{\bf Left:} Track-like events. {\bf Right:} Cascade-Like events.
In green the background from NC events is given. The blue histogram
shows the total background including misreconstructed CC events. 
Magenta histograms are for the total event rate (signal plus background) for NH (solid) and IH (dashed)
and various values of $\phi_{CP}$.
}\label{misreco}
\end{figure}

In a real detector it will not be possible to separate track-like and cascade-like events with 100\% efficiency
as discussed in Section~4.
The probability to misidentify a cascade as track $\epsilon(E_\nu)$ and the misidentification probability of 
a track as cascade $\eta(E_\nu)$ as introduced in Section~5 will be used now. 
With these two quantities the total background for the two event samples can be written
\begin{equation}
\begin{split}
\frac{dN_{bg}^{track}}{dE_\nu} = &\epsilon\frac{dN_{sig}^{casc}}{dE_\nu} 
+ \left[\epsilon(1-BR_{\tau\mu})+(1-\eta)BR_{\tau\mu}\right]\frac{dN_{\tau}}{dE_\nu}
+ \epsilon\frac{dN_{NC}}{dE_\nu}\\
\frac{dN_{bg}^{casc}}{dE_\nu} = &\eta\frac{dN_{sig}^{track}}{dE_\nu} 
+ \left[(1-\epsilon)(1-BR_{\tau\mu})+\eta BR_{\tau\mu}\right]\frac{dN_{\tau}}{dE_\nu}
+ (1-\epsilon)\frac{dN_{NC}}{dE_\nu}.
\end{split}
\label{eqmisreco}
\end{equation}
$BR_{\tau\mu}$ stands for the muonic branching ratio of the tau decay (17.4\%).
The total number of observed events in each channel is given by adding the (reduced) number of 
signal events and the background contribution from Equation~\ref{eqmisreco}:
\begin{equation}
\begin{split}
\frac{dN_{tot}^{track}}{dE_\nu} = &(1-\eta)\frac{dN_{sig}^{track}}{dE_\nu} 
+\frac{dN_{bg}^{track}}{dE_\nu}\\
\frac{dN_{tot}^{casc}}{dE_\nu} = &(1-\epsilon)\frac{dN_{sig}^{casc}}{dE_\nu} 
+\frac{dN_{bg}^{casc}}{dE_\nu}.
\end{split}
\label{tot}
\end{equation}
The resulting rates are shown in  
Figure~\ref{misreco} and quoted in Table~\ref{tabmisreco}.
\begin{table}[htpb]
\begin{center}
\begin{tabular}{||l|c|c|c|c||} \hline
Channel & Tracks NH & Tracks IH & Cascades NH & Cascades IH \\ \hline 
No oscil          & \multicolumn{2}{c|}{26315}     &  \multicolumn{2}{c||}{---} \\ \hline
Signal         &  8990  &  8735     &  1134-1547 & 350-519 \\ \hline
Misreco        & 232-329 & 47-79 &    1326 & 1280  \\
$\nu_\tau$        &  324-332  & 351-355  &  978-998 & 1057-1068  \\
NC                & 1092 & 1092   & 3640 & 3640   \\ \hline
BG Total       & 1655-1745 & 1494-1522 &  5944-5964 & 5977-5988 \\ \hline
Total          &  10645-10736 & 10229-10257  & 7099-7491 & 6338-6496 \\ \hline
\end{tabular}
\end{center}
\caption{Event numbers for $N_{pot}=1.5\cdot10^{21}$ in the track and cascade channel for both mass hierarchy schemes 
and varying $\phi_{CP}$ values.}
\label{tabmisreco}
\end{table}
The signal contributions are reduced by 15-20\%. The track-like sample is still dominated by signal. 
The different background channels add up to 15-18\% of the overall rate. 
The situation for the cascade-like events does not change very much compared to the numbers given in
Table~\ref{tabclean}. Some backgrounds are reduced due to migration into the track-like channel (e.g. NC), 
others are increased. The total number of background events is almost unaffected by the choice of MH 
(and by $\phi_{CP}$) as the MH dependence of different backgrounds has the tendency to cancel.
This leaves the event difference of the signal part almost unaffected. 
The event rates between the two MH hypotheses differ now by 9-18\% with a statistical uncertainty of 1.2\%.
The statistical significance of the MH hypothesis test is still better than 7$\sigma$
and it remains at the level of 3$\sigma$ even when adding
an additional systematic uncertainty of 3-4\% (depending on the true value of the CP phase)
for the determination of the total cascade event rate.
The knowledge of the detector performance, water parameters, neutrino cross sections, oscillation
parameters and the monitoring of the neutrino flux contribute to the systematic uncertainty. 
The neutrino flux normalisation can be controlled
by performing a complementary measurement of $\nu_\mu$ CC events. Uncertainties of the oscillation
parameters will have been reduced by ongoing experiments and neutrino cross sections will have been
measured with high precision by ongoing and planned short-baseline experiments by the time the here proposed
experiment is running.  The water in the abyss of the Mediterranean Sea is extensively studied in the
ANTARES experiment. All these measurements will help reducing the systematic uncertainty.

Figure~\ref{misreco} illustrates the total event numbers as function of neutrino energy, detailing the different
contributions. NC events are added to the $E_\nu/2$ bins according to their lower light yield in the detector.
Their main contribution is found for energies below 6~GeV. Background from CC events is 
instead mainly seen above 5~GeV. The separation of the MH hypotheses is most pronounced in the range 4-8 GeV.

\section{Conclusion}

A neutrino beam from IHEP Protvino to a Mton detector installed in the abyss of the Mediterranean Sea has a
baseline close to the ``magic" value of 2600~km. A low energy ``phase-1" part of the future KM3NeT neutrino
telescope could serve as target for such a beam. Counting of cascade-like events would allow a measurement of the
neutrino mass hierarchy with a significance of 3$\sigma$ for $1.5\cdot10^{21}$ protons on target and a
systematic uncertainty of 3-4\% for the event rate determination.
Higher values of the significance can be reached by reducing the systematic uncertainty,
considering a moderate capability to measure the neutrino energy of the 
signal events or an improved flavour tagging method.  
The proposed measurement would be complementary to an analysis of atmospheric neutrinos 
in the same detector. 
The combination of both measurements will possibly allow an unambiguous determination of the neutrino mass hierarchy.
 
 
\section{Acknowledgment}

I would like to thank P. Coyle (CPPM), R. Nahnhauer (DESY)  and Ch. Spiering (DESY) for inspiring discussions
and A. Meregaglia (IPHC) for technical support with the Globes package.

\end{document}